 \documentclass[9pt,prd,aps,amssymb, twocolumn, amsmath,tightenlines]{revtex4}
 \usepackage{graphics}
 \usepackage[pdftex]{graphicx}

  \usepackage[utf8]{inputenc}
\usepackage{amsmath}
\usepackage{amssymb}
\usepackage{booktabs}

 \usepackage{amsmath,amssymb,amsthm}

\begin{document}
\preprint{}

\title{Carath\'eodory II: The Geometry of Financial Irreversibility}

\author{ Bernhard K. Meister}
 \email{bernhard.k.meister@gmail.com}



\date{\today}




\begin{abstract}
\noindent 
In quantum mechanics and finance, numeraire invariance - the unobservability of absolute phase or price scale - fits with a projective and curved state space. This projective geometry has a measurable signature. For spin-one and higher spin systems, the Taylor expansion of directed distance contains a non-zero cubic term,  which induces a fundamental asymmetry under the exchange of states. 
The Second Law, the failure of Maxwell's demon,  and the limitations of sequential traders can all be reduced to this asymmetry.
 \end{abstract}
 
\maketitle
\section{Introduction: Carath\'eodory's Formulation of the Second Law}

\noindent 
Carath\'eodory's formulation of the Second Law\cite{Caratheodory1909} begins not with entropy, but with an observation about state space:  In every neighborhood of any equilibrium state, there exist states that cannot be reached by adiabatic processes. 

\noindent 
The statement is deceptively simple. It says nothing about numerical quantities, nothing about disorder, nothing about  the arrow of time. It merely asserts that the state space has a particular connectivity. Carath\'eodory then showed that for the `simple systems' he considered, this local inaccessibility implies the existence of a global foliation into entropy surfaces. 

\noindent 
Following Buchdahl\cite{Buchdahl1966}, we adopt a  relational point of view. The Second Law is not about continuous trajectories, but about the ordering of equilibrium states. The directed divergence $D(P|Q)$ is the natural tool - it quantifies the `cost' of  switching from $P$ to $Q$ without asking what happens in between\footnote{Similarly, in quantum mechanics the state of the system between measurements is not necessarily fully known.}. 
One may view this as defining an ordering: `$Q$ is more costly to reach from $P$ than the reverse'. The cubic term $T_{ijk}$ captures the leading asymmetry in this cost. For finite-resource observers, this asymmetry translates into an unavoidable geometric tax  that makes certain transitions effectively irreversible, even when no heat is exchanged with the environment.

\noindent 

\noindent 
This paper takes  the Second Law as given. We do not derive it. We ask a different question: What geometry of state space, and what constraints on observers, naturally accommodates and illuminates the Law  in the quantum finite-resource regime?
The answer: the cubic term in a Taylor expansion is not zero.

\noindent 

\noindent 
Unlike in the earlier paper\cite{Meister2026a}, where it was assumed that a classical system  had specific finite properties, here a more general  answer, appropriate to the quantum finite-resource regime  is studied. Numeraire invariance, a natural property for quantum systems, makes state space projective and curved. For spin-one or other higher spin systems, the Taylor expansion of directed distance contains a non-zero cubic term that induces a directional bias; unlike the quadratic metric, this third-order contribution is sensitive to the ordering of states. This term is a geometric fact, independent of how one observes. Under sequential measurement, its effects become unavoidable - a geometric tax that blocks Maxwell demons and drives the Second Law.  The Carath\'eodory's inaccessibility turns   quantitative through this asymmetry.

\noindent 
This paper operates at the intersection of three scales. At the microscopic level, numeraire invariance makes the state space projective and curved. At the macroscopic level, the Second Law is an observed regularity.  
The bridge between them is the mesoscopic scale, where observers with finite resources cannot perform  joint measurements  to circumvent the collective-local gap introduced in section IV. The cubic term in the directed divergence is always present; its effects, however, can be mitigated by sufficiently clever measurement strategies. For observers restricted to standard sequential measurements, such mitigation is impossible - the cubic term's accumulated effects become unavoidable, generating a geometric force and restricting   Maxwell's Demon. The collective-local gap is the consequence. 

\noindent 
The companion paper\cite{Meister2026a} explored a different micro-to-macro link: the PKD theorem shows how finite sufficient statistics force a flat exponential family at macroscopic scales. Here, we uncover a more general mechanism where curvature, not flatness, is the source of thermodynamic behavior - a fact that becomes important only when viewed through the mesoscopic lens of finite resources or restricted measurements. 
The bridge between them is the mesoscopic scale, where for example observers with finite resources perform sequential measurements. At this scale, the cubic term in the directed divergence cannot be ignored; its effects accumulate, generating the geometric work cost, blocking Maxwell's demon. The collective-local gap is a measurable signature of this mesoscopic regime.  


\noindent 
Section II covers how numeraire invariance relates to curvature, Section III reviews the Taylor expansion, Section IV introduces Veronese submanifolds and the collective-local gap, Section V covers geometric drift and Maxwell's  demon, Section VI applies the ideas to  finance, and then the paper ends with a short conclusion as Section VII.


\section{Numeraire Invariance Forces Curvature}
\noindent 
In quantum mechanics, the overall phase of a state vector is unobservable. The states $|\psi\rangle$ and $e^{\i \lambda|}\psi\rangle$ for any real $\lambda$ represent the same physical situation. This is  numeraire invariance - only the direction, or  ray, matters.

\noindent 
The space of all such rays is the  projective Hilbert space  $\mathbb{C}P^n$, where $n$ is one less than the Hilbert space dimension. 
Curvature is not an optional extra; it is a built-in property of the space of rays.

\noindent 
For a three-level system (spin-$1$), the state space is $\mathbb{C}P^2$, a more complicated curved manifold. Its curvature is non-zero. 
As we shall see, this curvature has thermodynamic consequences when the observer has finite resources and can only perform sequential measurements.

\noindent 
Thus numeraire invariance---the fact that only relative quantities are observable---forces the true state space to be projective and therefore intrinsically curved\footnote{ Quantum mechanics leads to complex projective space $\mathbb{C}P^n$ (because the phase is complex), while finance leads to real projective space $\mathbb{R}P^n$ (because price ratios are positive real numbers). 
This distinction does not affect the core argument: both spaces are curved, and the cubic term $T_{ijk}$ appears in both settings when one considers the appropriate directed divergence. The geometric mechanisms - curvature, holonomy - operate in both. It does affect the  numerical examples (e.g., the  collective-local gap of Section IV); the finance analog would not inherit these exact numbers but the structural fact that $T_{ijk}\neq0$ imposes a similar geometric tax. 
In what follows we refer simply to `projective space'.}. In the spin-$1$ case, we are dealing with $\mathbb{C}P^2$, and its curvature will leave a measurable signature: the cubic term in the Taylor expansion of the distance between states
\footnote{In financial markets, `absolute' prices are also unobservable; only exchange rates (`relative' prices) matter. If all prices double overnight, the real economic state is unchanged. Hence market states are rays in price space, and the natural state space is again projective. The same curvature appears, and as we will see, it underlies the impossibility of arbitrage for sequential traders.}.

\section{The Taylor Expansion: Why the Cubic Term Matters}
\noindent 
Let $D(P\|Q)$ be any smooth `directed divergence'\footnote{The term `directed distance' was used loosely above. We consider divergences, not metric distances, because divergences can have a non-vanishing cubic term - as  in our setting. 
}
 between two nearby states. This could be relative entropy or any other asymmetric distinguishability measure\footnote{In quantum mechanics, this could be the quantum relative entropy or the quantum Jensen–Shannon divergence. In finance, a directed divergence for classical return distributions is the Kullback–Leibler divergence.}; the specific choice does not matter for the argument.  Expanding around a reference state $P$, with $Q$ at an infinitesimal displacement $dP$, we have\footnote{In finance, the same expansion applies to the log return between two portfolios. The quadratic term captures volatility, while the cubic term captures skewness - the directional bias that makes round trips costly.}
\begin{equation}
D(P\|P+dP) = \frac{1}{2} g_{ij} dx^i dx^j + \frac{1}{6} T_{ijk} dx^i dx^j dx^k + \cdots .\nonumber
\end{equation}
\noindent 
The quadratic term defines a  tensor $g_{ij}$\footnote{  
The two canonical choices: For quantum systems, the quadratic term yields the Fubini-Study metric on $\mathbb{C}P^n$; for financial markets, it yields the Fisher-Rao metric on the statistical manifold of distributions.
The argument, however, requires only that  $g_{ij}$ is symmetric, not its specific form.} induced by the chosen divergence. 

\noindent 
The cubic term is different. The tensor $T_{ijk}$, called the Amari-Chentsov tensor\cite{amari2016} in information geometry in the case of relative entropy, generates the antisymmetric component  of the directed divergence
\begin{equation}
D(P\|P+dP) - D(P+dP\|P) = \frac{1}{3} T_{ijk} dx^i dx^j dx^k + \cdots\nonumber
\end{equation}
\noindent 
The quadratic terms cancel. The leading contribution to irreversibility is the cubic term. If $T_{ijk} = 0$, the world is reversible to this order; if $T_{ijk} \neq 0$, there is a built-in asymmetry: the displacement from $P$ to $Q$ is not the same as the displacement from $Q$ to $P$.

\noindent 
Most treatments stop at the quadratic approximation, implicitly assuming that cubic and higher terms are negligible. For spin-$1/2$ systems (the Bloch sphere), this is justified: the geometry is such that $T_{ijk}=0$. But for spin-$1$ systems, $T_{ijk} \neq 0$\footnote{This statement holds for the canonical directed divergences - relative entropy and its quantum generalizations. Metric distances (Bures, Hellinger) have $T_{ijk}=0$ for all systems by construction, but do not capture the asymmetry of finite-resource measurement costs.}. 
 The cubic term is the leading contribution to irreversibility - and under sequential measurement, its effects accumulate.

\noindent 
The Second Law, in this view,  finds its microscopic seed in $T_{ijk} \neq 0$  - a fact that, under finite-resource observations, grows into macroscopic irreversibility. The question `Why does entropy increase?' becomes `Why is the cubic term non-zero?' - and the answer lies in the geometry forced by numeraire invariance and the finite resources of the observer, as we will see in the next section.

\section{The Veronese Submanifold and the Collective-Local Gap}
\noindent 
The geometric structure discussed in this section extends beyond quantum mechanics to financial markets. If absolute prices don't matter, only ratios, then market states are rays in a projective space—curved for the same reason quantum state spaces are curved. Sequential trading (one asset at a time) traps one on a submanifold, and the cubic term appears as an unavoidable cost.

\noindent As an example, we consider a spin-$1$ quantum system.
Such a system can be realized as the symmetric subspace of two spin-$1/2$ particles~\cite{BrodyHughston}. The full state space is then $\mathbb{C}P^2$, but not all states are accessible to an observer with finite resources.
 
\noindent 
To see how finite resources restrict access, consider two limitations that confine the observer to product states. First, unitary evolution acting independently on each particle (the 
U-process, local unitaries). Second, sequential measurements (
R-process) that measures one particle at a time, projecting the state after each measurement. Both keep the state on  the submanifold of product states, and
the accessible states lie on the  Veronese submanifold  $\nu \subset \mathbb{C}P^2$ - the set of separable (non-entangled) states. 

\noindent 
The Veronese submanifold $\nu$ is itself curved, so parallel transport around a loop in $\nu$ accumulates a non-trivial holonomy - a geometric phase. This holonomy represents information that becomes inaccessible to the sequential observer  because their restricted measurements cannot resolve the geometric phase\footnote{
The Veronese submanifold $\nu$ is isometric to $\mathbb{C}P^1$, the ordinary Bloch sphere, which has constant positive curvature. A sequential observer confined to $\nu$ therefore experiences non-trivial holonomy when traversing a closed loop; this holonomy is the geometric origin of the Bagan residue.  In principle, an observer with access to the full unitary group or joint measurements could reverse the holonomy by leaving $\nu$, reversing each step. or using entangled operations. For a finite-resource observer restricted to $\nu$, however, it is irreversible. This is the operational meaning of `cannot be undone'.
}. 

\noindent 
\noindent 
For a spin-$1$ system under sequential observations, the gap between the information obtainable by sequential measurements and the information available from a hypothetical joint measurement is non-zero and positive.  Here we consider the difference in the optimal fidelity from adaptive sequential measurements and a collective measurement. 
This leads to a  collective-local gap~\cite{Bagan2006}. For spin-$1/2$ systems, the gap is zero; for higher spins it drops off and in the classical limit $s \to \infty$ it vanishes. 

\noindent 
The  formula for the collective-local gap
\footnote{
The collective-local gap exists across different sequential protocols (adaptive or not) and performance criteria.
For the gap calculation, we consider the difference between collective and sequential fidelities for systems of $N$ spin-$1/2$ copies - interpreted here as spin-$s$ states where $s=N/2$—yielding $F_{col}(N) = \frac{N+1}{N+2}$ and $F_{seq}(N) = \frac{N+1}{N+2} - \frac{1}{N(N+1)}$\cite{Bagan2006}. \\
\noindent
Furthermore, $T_{ijk}$ is a local geometric coefficient, whereas $F_{seq}$ and $F_{col}$ are integrated global results. 
The global fidelities respond to the full metric structure and manifold topology; the gap $1/N(N+1)$ constitutes the macroscopic signature of the microscopic/local asymmetry encoded in $T_{ijk}$.}
 as a function of total spin $s$ is 
\begin{itemize}
    \item $s = 1/2$: gap $= 0$ (seq. and coll. meas. coincide),
    \item $s = 1$: gap $= 1/6$,
        \item $s = 3/2$: gap $= 1/12$,
    \item $s \to \infty$: gap $\to 0$.
\end{itemize}

\noindent 
Thus the Veronese restriction - the inevitable consequence of finite resources - makes the cubic term of Section~III not just a mathematical possibility but an operational necessity. The  gap is the numerical signature of this geometry.

\section{The Geometric Reason Why Systems Drift}
\noindent 
For  this section, we confine the analysis to the canonical directed divergences: the quantum relative entropy in quantum mechanics and the Kullback–Leibler divergence in finance. For these choices, the cubic coefficient 
$T_{ijk}$  coincides (up to conventional scaling) with the Amari–Chentsov tensor of information geometry. 

\noindent

\noindent 
We now examine how the cubic term - the driver of this asymmetry - produces a directional bias that, for finite-resource observers, becomes an unavoidable geometric tax.

 
\subsection{ Work Surcharge: Ethereal Spheres\protect\footnote{In Mysterium Cosmographicum (1596) Kepler  tied the orbits of the planets to the Platonic solids; an idea he later relinquished to advance. } 
}
\noindent 
From Section~III, the directed divergence between two nearby states expands as
\begin{equation}
D(P\|P+dP) = \frac{1}{2}g_{ij}dx^i dx^j + \frac{1}{6}T_{ijk}dx^i dx^j dx^k + \cdots .\nonumber
\end{equation}
The quadratic term gives the metric; the cubic term $T_{ijk}$ is the Amari-Chentsov tensor and encodes the asymmetry. 

\noindent 

\noindent 


\noindent 
For an infinitesimal step $dx^i$, the additional work cost   beyond the reversible metric part - the quadratic term is even under reversal:  
$dx \rightarrow $ -$dx$ -  is the cubic term
\begin{equation}
dW_{\text{surcharge}} = \frac{1}{6}T_{ijk}dx^i dx^j dx^k .\nonumber
\end{equation}
\noindent 
The cubic asymmetry of the directed divergence creates a geometric tilt. The Second Law is the empirical record of this bias.
 
 \noindent 
\noindent 

\noindent 

\noindent 


\noindent 
\noindent 


\noindent 
The infinitesimal work surcharge $dW_{\text{surcharge}} = \frac{1}{6}T_{ijk}dx^i dx^j dx^k$ 
is a geometric fact: every step in a curved space carries this local asymmetry. But for a 
finite-resource sequential observer traversing a finite loop, these infinitesimal costs 
accumulate into an unavoidable tax. It is this accumulated tax  - not the local asymmetry itself - that makes the return trip 
operationally inaccessible\footnote{An infinite-resource observer, able to perform joint measurements 
and leave the submanifold $\nu$, can avoid the tax.  The cancellation requires, for example, the full unitary group on $\mathbb{C}P^2$ or entangled   operations, which are unavailable to an observer confined to $\nu$. This is the   operational content of the collective - local gap of Section~IV.}.
Next, we look at the Maxwell's demon - and try to understand why geometry acts as a block.
  
\subsection{Demon's Geometry Problem} 
\noindent 

\noindent 
Maxwell's demon attempts to move a system from a more mixed\footnote{Here `mixed' is used in the classical sense.}  state  to a more sorted state. The cubic term makes process costly, and the demon has to carry out geometric work 
\begin{equation}
W_{\text{demon}} = \sum_{\text{sorting steps}} \frac{1}{6}T_{ijk}dx^i dx^j dx^k \geq 0 .\nonumber
\end{equation}\noindent 
This is not about erasing memory or information; it is the cost of moving against the grain of the manifold. The demon faces resistance\footnote{This resistance provides an alternative geometric origin of Landauer's principle. The familiar $kT\ln2$ bound lives within the quadratic metric description; fluctuations can in principle affect this contribution (Norton's critique). The cubic term $T_{ijk}\neq0$, however, makes the backward shell volume structurally smaller than the forward shell volume;  imposing an additional asymmetry that fluctuations cannot erase.
The total work required is the sum of these two contributions, but only the cubic term is truly unavoidable for a finite-resource observer. The demon's work $W_{\text{demon}} = \sum \frac{1}{6}T_{ijk}\Delta x^i \Delta x^j \Delta x^k$ quantifies this   tax.
}. 
This is captured in the argument by Smoluchowski of the Brownian ratchet failure.

\noindent 

\noindent 


\subsection{  LPs Live Longer\protect\footnote{LPs, the acronym for Liquidity Providers,  are counterparties to LTs (liquidity takers),  market impact\cite{Meister2026b} is the consequence.}}
\noindent 
In finance, the same tax appears. The market maker's spread is the observable manifestation of this tax

\begin{equation}
\text{Spread} = \sum_{\text{typical trades}} \frac{1}{6}T_{ijk}dx^i dx^j dx^k \sim \frac{1}{6} \langle T_{ijk}\rangle .\nonumber
\end{equation}
\noindent 
where $\langle T_{ijk} \rangle$ denotes the   average over  round trips. This averaged quantity is not zero precisely because $T_{ijk} \neq 0$. 

\noindent 

\noindent 



\noindent 

\noindent 


\noindent 


\section{ Geometry of Markets}

\noindent 
Numeraire invariance makes financial state space projective and curved; sequential trading traps traders on submanifolds, just as in quantum mechanics; sequential trading can trap traders on a submanifold, such as a Gaussian one.

\noindent 
In a non-Gaussian market, returns are not fully described by drift and  variance alone. Higher-order cumulants—skewness, kurtosis, and beyond—are non-zero and relevant. In geometric terms, these signal curvature of the underlying statistical manifold: the Amari–Chentsov tensor $T_{ijk}$ is non-zero. The cubic term  is the leading signature of this curvature.

\noindent 
However, even in a curved space, there exist submanifolds where the divergence asymmetry vanishes ($T_{ijk}=0$). In finance, these are the Gaussian subspaces where all cumulants beyond the second vanish, and the Amari-Chentsov tensor is zero. In quantum mechanics, the analogous flat submanifolds (in the sense of divergence asymmetry) are the spin-$1/2$ state spaces: $\mathbb{C}P^1$ is metrically curved, but its directed divergence has no cubic term, so $T_{ijk}=0$. For the spin-$1$ system considered here, the Veronese submanifold $\nu$ consists of product states of two spin-$1/2$ particles in the symmetric subspace. While a single spin-$1/2$ has $T_{ijk}=0$, the embedding of $\nu$ in $\mathbb{C}P^2$ induces a non-zero cubic term, leading to the geometric tax.

\noindent
The essential point is that sequential traders cannot remain on these   submanifolds. 
In quantum mechanics, a sequential observer measuring one particle at a time remains on the Veronese submanifold $\nu$ (the set of product states). However, because $\nu$ itself is curved, the observer accumulates holonomy when traversing closed loops, leading to the geometric tax. In finance, by contrast, a sequential trader is a `price-taker', as  the market's state is given, not chosen. If the market is non-Gaussian, its state lies off the flat Gaussian submanifold, and the trader cannot force it back onto that submanifold. The geometric tax arises from the market's intrinsic curvature, not from the trader's path.

\noindent 
Consider a three-currency triangle (USD $\to$ EUR $\to$ GBP $\to$ USD). In log-space, the product of exchange rates becomes a sum. Let $x_{12}, x_{23}, x_{31}$ be the log-returns for each leg. Expanding each,
\begin{equation}
\log(1+x) = x - \frac{1}{2}x^2 + \frac{1}{3}x^3 + \cdots\nonumber
\end{equation}
The net log-return around the triangle is
\begin{equation}
x_{12} + x_{23} + x_{31} - \frac{1}{2}\bigl(x_{12}^2 + x_{23}^2 + x_{31}^2\bigr) + \frac{1}{3}\bigl(x_{12}^3 + x_{23}^3 + x_{31}^3\bigr) + \cdots\nonumber
\end{equation}

\noindent
If the market's state is confined to a flat submanifold\footnote{Here `flat' refers to vanishing divergence asymmetry  ($T_{ijk}=0$), not metric flatness. For example, the Bloch sphere $\mathbb{C}P^1$ is metrically curved but has $T_{ijk}=0$; Gaussian subspaces are both. The relevant condition for the geometric tax is $T_{ijk}\neq0$, regardless of metric curvature.}, the cubic sum $  \sum x_i^3$ averages to zero. No systematic bias appears.


\noindent

\noindent
But because sequential trading in a curved market inevitably explores directions where $T_{ijk}\neq0$, the geometry imposes a non-zero expectation for the cubic contribution. This means   the trader faces a systematic expected loss on any circular trade - a geometric tax. The cubic term itself is harmless on flat submanifolds; it is the combination of non-zero curvature and sequential access that makes it treacherous.

\noindent
Thus the same cubic term that signals curvature in quantum thermodynamics also, under sequential measurement that explores curved directions, guarantees that a sequential trader cannot profit from a circular trade. The outcome is either zero (in the flat case) or a loss (in the curved case) - never a gain. The Second Law and the `no-profit' condition share the same geometric origin: $T_{ijk}\neq0$.



\noindent 

\noindent 
For sequential traders confined to curved regions of projective price space, the cubic term $T_{ijk}$ (skewness) generates a systematic loss on circular trades\footnote{A trader with infinite resources (joint measurements, full unitary control) could in principle profit from or cancel the cubic term, but such traders are idealizations. The axiom applies to the realistic case: finite-resource sequential observers.} - a geometric tax. This tax is not merely an asymmetry; the Second Law\footnote{Numeraire invariance plus sequential constraint yields asymmetry ($D(P\|Q)\neq D(Q\|P)$). The Second Law supplies the arrow: the costly direction is the one the trader traverses. Without this directional postulate, one could imagine a world where sequential traders consistently profit - a perpetual motion machine of arbitrage. The Second Law forbids this.} also supplies the direction: it always acts against the trader. The tax vanishes on flat submanifolds ($T_{ijk}=0$) and persists whenever third moments exist, regardless of jumps or non-stationarity.

\noindent 

\noindent 

\noindent 

\noindent 



\section{Conclusion}
\noindent 
The Second Law has been interpreted  as a statement about disorder, about missing information, and about the arrow of time. It is, rather, a geometric fact: the cubic term of the Taylor expansion of the directed divergence is non-zero.
Quantum thermodynamics, finance, and the classic puzzle  of Maxwell's demon  can be  examined through this lens. In the quadratic approximation, the world appears reversible. Distances are symmetric; closed loops close perfectly; entropy is a state function. This is the spin-$1/2$ world,   classical thermodynamics with infinite resources,  or flat markets with no arbitrage. It is a useful approximation, but it jars with our experience. In the classical limit, where the cubic term vanishes, the Pitman–Koopman–Darmois theorem offers a way to rescue the Second Law, as shown in the companion paper\cite{Meister2026a}.

\noindent 
The cubic term in spin-one and higher spin  systems is direction dependent. Every round trip pays a geometric `tax'. In thermodynamics this tax is entropy; in finance it can be the liquidity provider's revenue. The tax restricts Maxwell's demon. 

\noindent 
At the microscopic level, the fundamental laws   are reversible in the sense that the quadratic  approximation dominates the directed divergence expansion. Only observer with unlimited resources - someone who can perform all joint measurements, implement any unitary operation, and track every degree of freedom - can  saturate the fundamental estimation limit that the full expansion - including the cubic term - imposes.  Non-idealised observers have finite resources. They cannot, as a consequence, compensate for the asymmetry of the next order term, the cubic term, which gains importance  at the mesoscopic scale. 
%
For them, it becomes an operational hurdle. Its effects cannot be averaged away or circumvented; the result is an unavoidable  cost - the collective-local gap - paid on top of the fundamental limit.
In this view, the Second Law is not a property of the world in isolation, but of the relationship between the world and any  finite-resource observer\footnote{Imagine an egg falling and its shell cracking on the floor. The cubic term
is the geometric fact: the floor. Finite resources: the egg's fragility. The `tax' - the breaking - requires both. The external finite-resource observer cannot put the egg back together again. 
}. The cubic term is the lowest order mathematical signature  of this coarse-graining. 
The geometric argument   does not merely restate the Second Law  -  it arguably reveals its origin. 
 \\
  \\
 \noindent
 {\it Ein Teil von jener Kraft, die stets das B\"{o}se will und stets das Gute schafft.}
   \,\,\, \,\,\,\,\,    \,  \,\,   \,\, Goethe, Faust I\footnote{The cubic term is Mephistophelean: 
   `Part of that force which eternally wills evil and eternally works good'. It aims to to increase disorder, to lose information, to block returns and to impose a geometric tax, but always creates good: the Second Law, the arrow of time, and viable markets.}

\noindent 

\end{document}